\documentclass[prc,aps,amsfonts,amsmath,superscriptaddress,showpacs,floatfix,onecolumn]{revtex4-1}
\usepackage{graphicx} 
\usepackage{epsfig} 
\usepackage{rotating}
\usepackage{subfig}
\usepackage{color}

\begin{document}

\title{Spin-orbit interaction in relativistic nuclear structure models}

\author{J.-P. Ebran}
\affiliation{CEA,DAM,DIF, F-91297 Arpajon, France}
\author{A. Mutschler}
\affiliation{Institut de Physique Nucl\'eaire, Universit\'e Paris-Sud, IN2P3-CNRS, 
F-91406 Orsay Cedex, France}
\author{E. Khan}
\affiliation{Institut de Physique Nucl\'eaire, Universit\'e Paris-Sud, IN2P3-CNRS, 
F-91406 Orsay Cedex, France}
\author{D. Vretenar}
\affiliation{Physics Department, Faculty of Science, University of
Zagreb, 10000 Zagreb, Croatia}

\begin{abstract} 

Relativistic self-consistent mean-field (SCMF) models naturally
account for the coupling of the nucleon spin to its orbital motion,
whereas non-relativistic SCMF methods necessitate a phenomenological
ansatz for the effective spin-orbit potential. Recent experimental
studies aim to explore the isospin properties of the effective
spin-orbit interaction in nuclei. SCMF models are very useful in the
interpretation of the corresponding data, however standard 
relativistic mean-field and non-relativistic Hartree-Fock models use
effective spin-orbit potentials with different isovector properties,
mainly because exchange contributions are not treated explicitly in
the former. The impact of exchange terms on the effective spin-orbit 
potential in relativistic mean-field models is analysed, and it is
shown that it leads to an isovector structure similar to the one used
in standard non-relativistic Hartree-Fock.
Data on the isospin dependence of spin-orbit splittings in spherical
nuclei could be used to constrain the isovector-scalar channel of
relativistic mean-field models. The reproduction
of the empirical kink in the isotope shifts of even Pb nuclei by
relativistic effective interactions points to the occurrence of
pseudospin symmetry in the single-neutron spectra in these nuclei.

\end{abstract}

\pacs{21.10.-k,21.60.Jz}

\date{\today}

\maketitle

\section{Introduction}

Self-consistent mean-field models based on relativistic energy density functionals (EDFs) 
with density-dependent strength parameters \cite{dd}, have been successfully applied 
to studies of a broad variety of nuclear phenomena such as radii, masses,
collective modes, fission and shape coexistence (see e.g.
\cite{Ring96,rprc}). Remarkable results have been obtained both in the
relativistic mean-field (RMF) framework \cite{Ring96,Niksic_2002}, and
more recently using the relativistic Hartree-Fock (RHF) scheme 
\cite{par_PKO2,ebr11,gu}, even though RHF applications have mostly been 
restricted to spherical nuclei. One of the basic advantages of using functionals 
with manifest covariance is the natural inclusion of the nucleon spin degree 
of freedom, and the resulting nuclear spin-orbit potential which emerges automatically 
with the empirical strength in a covariant formulation \cite{FS.00}. 

Non-relativistic EDFs such as, for instance, Skyrme \cite{Vautherin72} or the
Gogny \cite{berg} functionals must, of course, also include a spin-orbit term. 
In this case, however, the strength of the phenomenological spin-orbit term has 
to be adjusted to data on the energy spacing between spin-orbit partner states. 
This approach has been extensively applied and refined over the last four 
decades \cite{ben,SR.07,EKR.11}, and it provides an effective description of 
spin-orbit effects in nuclei.

The omission of an explicit treatment of exchange terms in the RMF
approach may have an impact on the description of the isovector
channel, in particular for the energy gap between spin-orbit partner
states when the ratio between neutrons and protons becomes very large.
The modification of spin-orbit splittings predicted by RMF-based
models differs from that obtained with non-relativistic, e.g. Skyrme
models \cite{Ring96,shar}.  Empirical constraints can be obtained by
studying the changes in neutron spin-orbit splittings when the number
of protons change and vice versa, as in the recent study of
spectroscopic properties of $^{35}$Si and $^{37}$S \cite{burg}. This
task is, however, not straightforward because single-particles
energies and occupations factors are not direct observables
\cite{duguet}.

On the theoretical side, the difference between the isospin dependence
of RMF and non-relativistic spin-orbit interactions can be analyzed by
performing a non-relativistic reduction of the Dirac equation. Such a
study was reported, for instance, by Sulaksono et al. \cite{burn} with
the goal to compare in a global way the magnitude of spin-orbit terms
in these two approaches. In this work we focus on the isospin
dependence of the spin-orbit effect using relativistic EDFs with
density-dependent strength parameters, and evaluate the effect of
explicit treatment of exchange terms in relativistic structure models.

\section{Spin-orbit term in relativistic effective interactions}

\subsection{\label{RMF}The RMF case}

Most SCMF models based on the relativistic mean-field approximation
have been formulated using the finite-range meson-exchange
representation, in which the nucleus is described as a system of Dirac
nucleons coupled to mesons fields through an effective Lagrangian. The
isoscalar scalar $\sigma$-meson, the isoscalar vector $\omega$-meson,
and the isovector vector $\rho$-meson build the minimal set of meson
fields that, together with the electromagnetic field, is necessary for
a description of bulk and single-particle nuclear properties. The
corresponding Lagrangian density reads
\begin{eqnarray}
\mathcal{L} =&& \bar \Psi (i \gamma^\mu \partial_\mu - m)\Psi \nonumber \\
&&+ \frac{1}{2}(\partial_\mu \sigma \partial^\mu \sigma - m_\sigma^2 \sigma^2) - \frac{1}{2}(\frac{1}{2}\Omega^{\mu\nu}\Omega_{\mu\nu} - m_\omega^2\omega_\mu \omega^\mu) - \frac{1}{2}(\frac{1}{2} \boldsymbol{R}_{\mu\nu} \cdot \boldsymbol{R}^{\mu\nu} - m_\rho^2 \boldsymbol{\rho}_\mu \cdot \boldsymbol{\rho}^\mu) \nonumber \\
&&-g_\sigma(\rho_B (\vec r))\bar \Psi \sigma \Psi -g_\omega(\rho_B (\vec r))\bar \Psi \gamma_\mu \omega^\mu \Psi -g_\rho(\rho_B (\vec r))\bar \Psi \gamma_\mu \boldsymbol{\rho}^\mu \cdot \boldsymbol{\tau} \Psi, 
\end{eqnarray}
where, for simplicity, we omit the Coulomb term which is not relevant for the present discussion. 
$\Psi$ denotes the nucleon spinor, $m$ is the nucleon bare mass, $m_\sigma$, $m_\omega$ and $m_\rho$ denote the meson masses. 
$\Omega^{\mu\nu}\equiv \partial^\mu\omega^\nu-\partial^\nu\omega^\mu$ and $\boldsymbol{R}^{\mu\nu}\equiv \partial^\mu\boldsymbol{\rho}^\nu-\partial^\nu\boldsymbol{\rho}^\mu$ are the $\omega$ and $\rho$ meson field tensors. Boldface symbols denote vectors and tensors in 
isospin space. The meson-nucleon couplings are assumed to be functions of 
the nucleon density (time-like component of the nucleon 4-current) $\rho_B(\vec r)$, and this density dependence 
in principle encodes all in-medium many-body correlations. 
\newline

In the self-consistent RMF framework the dynamics of independent nucleons is determined by 
local scalar and vector self-energies.  For simplicity spherical nuclei are considered and time-reversal symmetry is assumed 
(pairwise occupied states with Kramers degeneracy), which 
ensures that the only non-vanishing components of the 4-vector fields are the time-like 
ones and thus there is no net contribution from nucleon currents. Because of charge conservation 
only the 3rd component of the vectors in isospin space gives a non-vanishing contribution. 
The single-nucleon equation of motion is then the Dirac equation: 
\begin{equation}
\label{Dirac}
\left [ {\vec \alpha} \cdot {\vec p} + V + \beta (m+S) \right ] \psi_i = E_i \psi_i
\end{equation}
where $\vec \alpha = \gamma_0\vec\gamma$, $\beta = \gamma_0$,
$\gamma_0$ and $\vec\gamma$ are the Dirac matrices in Dirac representation,
and $\psi_i$ denotes the self-consistent solution for the i-th Dirac state of 
energy E$_i$: 
\begin{equation}
\left(
\begin{array}[c]{c}
\phi_i\\
\chi_i
\end{array}
\right)
\end{equation} 
with $\phi_i$ and $\chi_i$ denoting the large and small component, respectively. The scalar and time-like vector self-energies read:
\begin{equation}
\label{Self-enS}
S(\vec r) = g_\sigma(\rho_B(\vec r)) \sigma(\vec r)
\end{equation}

\begin{eqnarray}
V(\vec r) = && g_\omega(\rho_B(\vec r)) \omega(\vec r) + g_\rho(\rho_B(\vec r)) \tau^3 \rho(\vec r) + 
\frac{dg_\sigma}{d\rho_B}\sum_i\bar{\psi}_i(\vec r)\sigma(\vec r)\psi_i(\vec r) \\ 
&& + \frac{dg_\omega}{d\rho_B}\sum_i\bar{\psi}_i(\vec r) \gamma_0\omega(\vec r)\psi_i(\vec r)+ 
\frac{dg_\rho}{d\rho_B}\sum_i\bar{\psi}_i(\vec r)\gamma_0\rho(\vec
r)\tau^3\psi_i(\vec r) 
\label{Self-enV}
\end{eqnarray}

The explicit dependence of the coupling functions on the baryon
density $\rho_B$ produces rearrangement contributions to the vector
nucleon self-energy. The rearrangement terms result from the variation
of the couplings with respect to the baryon density.

In applications to nuclear matter and finite nuclei, relativistic
models are used in the {\em no-sea} approximation: the Dirac sea of
states with negative energies does not contribute to the densities and
currents.  In the nuclear ground state A nucleons occupy the lowest 
single-nucleon orbitals, determined self-consistently by the iterative
solution of the Dirac equation (\ref{Dirac}). Expressing the
single-nucleon energy as $E_i = m + \varepsilon_i$, where $m$ is the
nucleon mass, and rewriting the Dirac equation as a system of two
equations for $\phi_i$ and $\chi_i$, then, noticing that for bound
states $\varepsilon_i << m$, the equation for the upper component $\phi_i$ of the Dirac spinor reduces to the 
Schr\" odinger-like form \cite{bib82,rei89,lala} 
\begin{equation}
\label{Sch}
\left [ {\vec p} {1 \over 2 {\cal M}(r)} {\vec p} + U(r) +  V_{so}(r) \right ] \phi_i = \varepsilon_i   \phi_i
\end{equation} 
for a nucleon with effective mass

\begin{equation}
{\cal M}({r})\equiv m+\frac{1}{2}\left(S({r})-V({r})\right) \;,
\label{eq:meff}
\end{equation}

in the potential
$U(r) \equiv V(r) + S(r)$. The resulting additional spin-orbit potential \cite{bib82,rei89,lala}
\begin{equation}
\label{eq:vso}
V_{so} = \frac{1}{2r {\cal M}^2 (r)} \frac{\mathrm{d}}{\mathrm{d}r}(V-S)~\vec l \cdot \vec s 
\end{equation}
plays a crucial role in reproducing the empirical nuclear magic numbers.  
The non-relativistic limit corresponds to an $\frac{1}{2 {\cal M} (\vec r)}$ expansion. In the lowest order the isoscalar density $\rho_s(\vec r)$ can be approximated by the non-relativistic nucleon density $\rho_B(\vec r)$. 
At the energy scale characteristic for nuclear binding, meson exchange 
($\sigma$, $\omega$, $\rho$, $\ldots$) is just a
convenient representation of the effective nuclear interaction. The exchange
of heavy mesons is associated with short-distance dynamics that cannot be
resolved at low energies, and therefore in each channel meson exchange can be
replaced by the corresponding local four-point (contact) interactions between
nucleons. The relation between the two representations: finite-range
(meson exchange) and zero-range (point-coupling), is straightforward in nuclear
matter because of constant nucleon scalar and vector densities.
The Klein-Gordon equations of the
meson-exchange model with meson masses $m_{\phi}$ and
density-dependent couplings $g_{\phi}(\rho)$, are replaced by the corresponding
point-coupling interaction terms with strength parameters
$g_{\phi}^{2}/m_{\phi}^{2}$. In finite nuclei, however, because of 
the radial dependence of the densities, the expansion of the
meson propagator in terms of $1/m_{\phi}^{2}$ leads to a series of
gradient terms \cite{NVLR.08}. For the purpose of our discussion it suffices to 
consider only the lowest order, in which the 
self-energies $S(\vec r)$~(\ref{Self-enS}) and $V(\vec r)$~(\ref{Self-enV}) read
\begin{eqnarray}
S(\vec r) &=& - \frac{g_\sigma^2(\rho_B(\vec r))}{m_\sigma^2}\rho_B(\vec r) \label{eq:se0} \\
V(\vec r) &=& \frac{g_\omega^2}{m_\omega^2}\rho_B(\vec r) + \tau^3 \frac{g_\rho^2}{m_\rho^2}\rho_\tau(\vec r) \nonumber \\
&&- \frac{g_\sigma g_\sigma'}{m_\sigma^2} \rho_B^2(\vec r) + \frac{g_\omega g_\omega'}{m_\omega^2} \rho_B^2(\vec r) + \frac{g_\rho g_\rho'}{m_\rho^2} \rho_\tau^2(\vec r) \;,
\label{eq:se}
\end{eqnarray}
where $\rho_B$ is the nucleon density and $\rho_\tau$=$\rho^{(n)}_B-\rho^{(p)}_B$ is the isovector nucleon density. Introducing the notation: 
\begin{eqnarray}
\label{eq:def_alpha}
\alpha_i &\equiv& \frac{g_i^2}{m_i^2} \\
\alpha_i' &\equiv& \frac{\mathrm{d} \alpha_i}{\mathrm{d} \rho_B} = \frac{2g_i g_i'}{m_i^2} \\
\alpha_i'' &\equiv& \frac{\mathrm{d}^2 \alpha_i}{\mathrm{d} \rho_B^2} = 2\frac{(g_i')^2 + g_i g_i''}{m_i^2}
\end{eqnarray}
for  $i = \{\sigma,\omega,\rho\}$, and explicitly writing the neutron and proton contributions ($q = \{n,p\}$) with 
$\rho_B^{(q-q')} \equiv \rho_B^{(q)} - \rho_B^{(q')}$,
from Eqs (\ref{eq:vso}), (\ref{eq:se0}) and (\ref{eq:se}) one derives:
\begin{eqnarray}
\label{eq:Vso_W1W2} &&V_{so}^{(q)} = \frac{\alpha_\sigma +
\alpha_\omega + \alpha_\rho + 2 \alpha_\omega' \rho_B + 2 \alpha_\rho'
\rho_B^{(q-q')} + \frac{- \alpha_\sigma'' +
\alpha_\omega''}{2}\rho_B^2 + \frac{\alpha_\rho''}{2}
(\rho_B^{(q-q')})^2}{2r\left\{ m - \frac{1}{2}\left[ (\alpha_\sigma +
\alpha_\omega)\rho_B + \alpha_\rho \rho_B^{(q-q')} +
\frac{-\alpha_\sigma' + \alpha_\omega'}{2}\rho_B^2 +
\frac{\alpha_\rho'}{2}(\rho_B^{(q-q')})^2
\right]\right\}^2}\frac{\mathrm{d}\rho_B^{(q)}}{\mathrm{d}r} \vec l
\cdot \vec s \nonumber \\ &&+ \frac{\alpha_\sigma + \alpha_\omega -
\alpha_\rho + 2 \alpha_\omega' \rho_B + \frac{- \alpha_\sigma'' +
\alpha_\omega''}{2}\rho_B^2 + \frac{\alpha_\rho''}{2}
(\rho_B^{(q-q')})^2}{2r\left\{ m - \frac{1}{2}\left[ (\alpha_\sigma +
\alpha_\omega)\rho_B + \alpha_\rho \rho_B^{(q-q')} +
\frac{-\alpha_\sigma' + \alpha_\omega'}{2}\rho_B^2 +
\frac{\alpha_\rho'}{2}(\rho_B^{(q-q')})^2
\right]\right\}^2}\frac{\mathrm{d}\rho_B^{(q'\ne q)}}{\mathrm{d}r}
\vec l \cdot \vec s \nonumber \\
\end{eqnarray}
This expression can be rewritten as
\begin{equation}
\label{W1andW2}
V_{so}^{(q)} = \left[ W_1 \frac{\mathrm{d}\rho_B^{(q)}}{\mathrm{d}r} + W_2 \frac{\mathrm{d}\rho_B^{(q'\ne q)}}{\mathrm{d}r} \right]\vec l \cdot \vec s
\end{equation}
and the relevant ratio that determines the isospin dependence of the spin-orbit potential reads
\begin{equation}
\label{eq:W1W2}
\frac{W_1}{W_2}^{(q)}(\alpha_\sigma,\alpha_\omega,\alpha_\rho) \equiv
\frac{A^q(\alpha_\sigma,\alpha_\omega,\alpha_\rho)+B^q(\alpha_\rho,\rho_B^{(q-q')})}{A^q(\alpha_\sigma,\alpha_\omega,\alpha_\rho)-B^q(\alpha_\rho,0)}\;,
\end{equation}
\vspace{0.5cm}
with
\begin{equation}
A^q(\alpha_\sigma,\alpha_\omega,\alpha_\rho)\equiv
\alpha_\sigma + \alpha_\omega + 2 \alpha_\omega'
\rho_B + \frac{- \alpha_\sigma'' +
\alpha_\omega''}{2}\rho_B^2 + \frac{\alpha_\rho''}{2}
(\rho_B^{(q-q')})^2
\end{equation}
and
\begin{equation}
B^q(\alpha_\rho,\rho_B^{(q-q')})\equiv
\alpha_\rho+ 2 \alpha_\rho' \rho_B^{(q-q')}
\label{eq:isov}
\end{equation}
Equation (\ref{eq:W1W2}) shows that the ratio W$_1$/W$_2$ differs from unity
because of the isovector contribution (\ref{eq:isov}), and is larger than one 
for B$^q> 0$.

\subsection{\label{RHF}The RHF case}

In the relativistic Hartree-Fock case, in which exchange terms are treated explicitly, because of non-locality it is not possible to derive a
simple analytic expression for the non-relativistic spin-orbit potential. For a direct comparison with the RMF case, one can first consider the point-coupling approximation to the meson-exchange RHF Lagrangian, and further perform a Fierz transformation to obtain a corresponding RMF Lagrangian \cite{liang}. The interacting part of the RHF Lagrangian reads:
\begin{equation}
\label{eq:Lsimp}
\mathcal{L}_{int} = -g_\sigma(\rho_B) \bar{\Psi} \sigma \Psi - g_\omega(\rho_B) \bar{\Psi} \gamma_\mu \omega^\mu \Psi - g_\rho(\rho_B) \bar{\Psi} \gamma_\mu \boldsymbol{\rho}^\mu \cdot \boldsymbol{\tau}\Psi
\end{equation}
In the lowest-order point-coupling approximation \cite{ddpc} the mesons fields can be expressed 
\begin{eqnarray}
\label{eq:champPC}
\sigma &=& - \frac{g_\sigma(\rho_B(\vec r))}{m_\sigma^2}\bar \Psi \Psi \nonumber \\
\omega^\mu &=&  \frac{g_\omega(\rho_B(\vec r))}{m_\omega^2}\bar \Psi \gamma^\mu \Psi \nonumber \\
\boldsymbol{\rho}^\mu &=&  \frac{g_\rho(\rho_B(\vec r))}{m_\rho^2}\bar \Psi \gamma^\mu \boldsymbol{\tau} \Psi \;,
\end{eqnarray}
and for the equivalent Lagrangian in the point-coupling approximation
\begin{equation}
\label{eq:lagdensity}
\mathcal{L}_{int}^{PC} = -\frac{1}{2}\alpha_\sigma(\bar \Psi \Psi)(\bar \Psi \Psi) -\frac{1}{2}\alpha_\omega(\bar \Psi \gamma_\mu \Psi)(\bar \Psi \gamma^\mu \Psi) -\frac{1}{2}\alpha_\rho(\bar \Psi \gamma_\mu \boldsymbol{\tau} \Psi)\cdot (\bar \Psi \gamma^\mu \boldsymbol{\tau} \Psi)
\end{equation}
one obtains the ground-state expectation value
\begin{equation}
\left\langle\mathcal{L}_{int}^{PC} \right\rangle = -\frac{1}{2} \alpha_\sigma\rho_s^2 -\frac{1}{2} \alpha_\omega\rho_B^2 -\frac{1}{2} \alpha_\rho\rho_{\tau}^2  +\frac{1}{2} \alpha_\sigma\rho_{s,exch}^2 +\frac{1}{2} \alpha_\omega\rho_{v,exch}^2 +\frac{1}{2} \alpha_\rho\rho_{\tau,exch}^2 \;.
\end{equation}

Using the Fierz transformation, the couplings $\tilde \alpha_i$ of the corresponding RMF Lagrangian are expressed in terms of 
those of $\mathcal{L}_{int}^{(PC)}$  \cite{liang}:
\begin{eqnarray}
\tilde \alpha_S &=& \frac{7}{8} \alpha_\sigma + \frac{1}{2} \alpha_\omega + \frac{3}{2}\alpha_\rho \\
\tilde \alpha_V &=& \frac{1}{8} \alpha_\sigma + \frac{5}{4} \alpha_\omega + \frac{3}{4}\alpha_\rho \\
\tilde \alpha_{tV} &=& \frac{1}{8} \alpha_\sigma + \frac{1}{4} \alpha_\omega + \frac{3}{4}\alpha_\rho \\
\tilde \alpha_{tS} &=& -\frac{1}{8} \alpha_\sigma + \frac{1}{2} \alpha_\omega - \frac{1}{2}\alpha_\rho \;.
\end{eqnarray}
The resulting Fierz Lagrangian:
\begin{equation}
\left\langle\mathcal{L}_{int}^{(Fierz)}\right\rangle = -\frac{1}{2} \tilde \alpha_S\rho_s^2
-\frac{1}{2} \tilde \alpha_V\rho_B^2 -\frac{1}{2} \tilde
\alpha_{tV}\rho_{\tau}^2 -\frac{1}{2} \tilde \alpha_{tS}\rho_{tS}^2
\label{lag}
\end{equation}
is then equivalent to the RMF Lagrangian of the previous section but,
in addition, an isovector scalar term appears because of the Fierz
transformation. Additional terms in the pseudoscalar and pseudovector
channels do not contribute to the self-consistent ground-state
solution. Using this expression in (\ref{eq:W1W2}) yields the ratio
$\frac{W_1}{W_2}^{(q)}$ for the Fierz Lagrangian: 

\begin{equation}
\label{eq:W1W2fierz0}
\frac{W_1}{W_2}^{(q)} \equiv \frac{W_1}{W_2}^{(q)}(\tilde \alpha_S,\tilde \alpha_V,\tilde
\alpha_{tV}+\tilde \alpha_{tS}) 
\end{equation}
with the explicit functional dependence
\begin{eqnarray}
\label{eq:W1W2fierz}
\frac{W_1}{W_2}^{(q)} = \frac{\tilde \alpha_S + \tilde \alpha_V + \tilde \alpha_{tV} + \tilde \alpha_{tS} + 2 \tilde \alpha_V' \rho_B + 2(\tilde \alpha_{tV}' + \tilde \alpha_{tS}')\rho_B^{(q-q')}}{\tilde \alpha_S + \tilde \alpha_V - \tilde \alpha_{tV} - \tilde \alpha_{tS} + 2 \tilde \alpha_V' \rho_B  + \frac{-\tilde \alpha_S'' + \tilde \alpha_V''}{2}\rho_B^2 + \frac{\tilde \alpha_{tV}'' + \tilde \alpha_{tS}''}{2}(\rho_B^{(q-q')})^2}\nonumber \\
+ \frac{\frac{-\tilde \alpha_S'' + \tilde \alpha_V''}{2}\rho_B^2 + \frac{\tilde \alpha_{tV}'' + \tilde \alpha_{tS}''}{2}(\rho_B^{(q-q')})^2}{\tilde \alpha_S + \tilde \alpha_V - \tilde \alpha_{tV} - \tilde \alpha_{tS} + 2 \tilde \alpha_V' \rho_B  + \frac{-\tilde \alpha_S'' + \tilde \alpha_V''}{2}\rho_B^2 + \frac{\tilde \alpha_{tV}'' + \tilde \alpha_{tS}''}{2}(\rho_B^{(q-q')})^2} \;.
\end{eqnarray}

The structure of equation (\ref{eq:W1W2fierz}) is similar to that of
Eq. (\ref{eq:W1W2}) but, in addition to the isovector-vector, it
contains also an isovector-scalar contribution but the strength
parameter $\tilde \alpha_{tS}$ of this channel is not independent. In
the meson-exchange representation this channel corresponds to the
exchange of a $\delta$-meson. The isovector-scalar meson $\delta$ can
be, of course, explicitly included in the model Lagrangian but, as it
has been often argued in the literature, it is difficult to determine
its coupling strength from available data on finite nuclei. In the RMF
meson-exchange model DD-ME$\delta$ developed and tested in
Ref~\cite{RM.11}, for instance, the isovector effective mass $m^*_p -
m^*_n$ derived from relativistic Brueckner theory was used to
determine the coupling strength of the $\delta$-meson and its density
dependence. It was noted, however, that the explicit inclusion of the 
isovector-scalar meson does not improve the accuracy of calculated
properties of finite nuclei such as masses and radii.

\section{Results and discussion}

Conventional non-relativistic Hartree-Fock mean-field calculations
based on the Skyrme or Gogny force use a spin-orbit potential without
explicit isospin dependence, and with a constant strength parameter.
The explicit treatment of the exchange term constraints the ratio of
the resulting constants in the expression of Eq.~(\ref{W1andW2}) to
$W_1/W_2 = 2$ \cite{ben,shar}. In some cases this choice is too
restrictive, but it can be relaxed if the effective interaction
is interpreted as resulting from an energy density functional in the
sense of Kohn-Sham density functional theory \cite{SR.07,EKR.11}. In
the relativistic mean-field approximation (cf. Sec. \ref{RMF}) a weak
isospin dependence of the effective spin-orbit potential arises
because of the $\rho$-meson contribution (in meson-exchange models) or
the isovector-vector term of the interaction Lagrangian (in
point-coupling models). Exchange terms are not computed explicitly
and, because of the way the spin-orbit potential Eq.~(\ref{W1andW2})
emerges in the non-relativistic reduction of the single-nucleon Dirac
equation, the ratio $W_1/W_2$ in Eq. (\ref{eq:W1W2}) explicitly
depends on proton and neutron densities. 

\begin{figure}[]  
      {\includegraphics[width=0.8\textwidth]{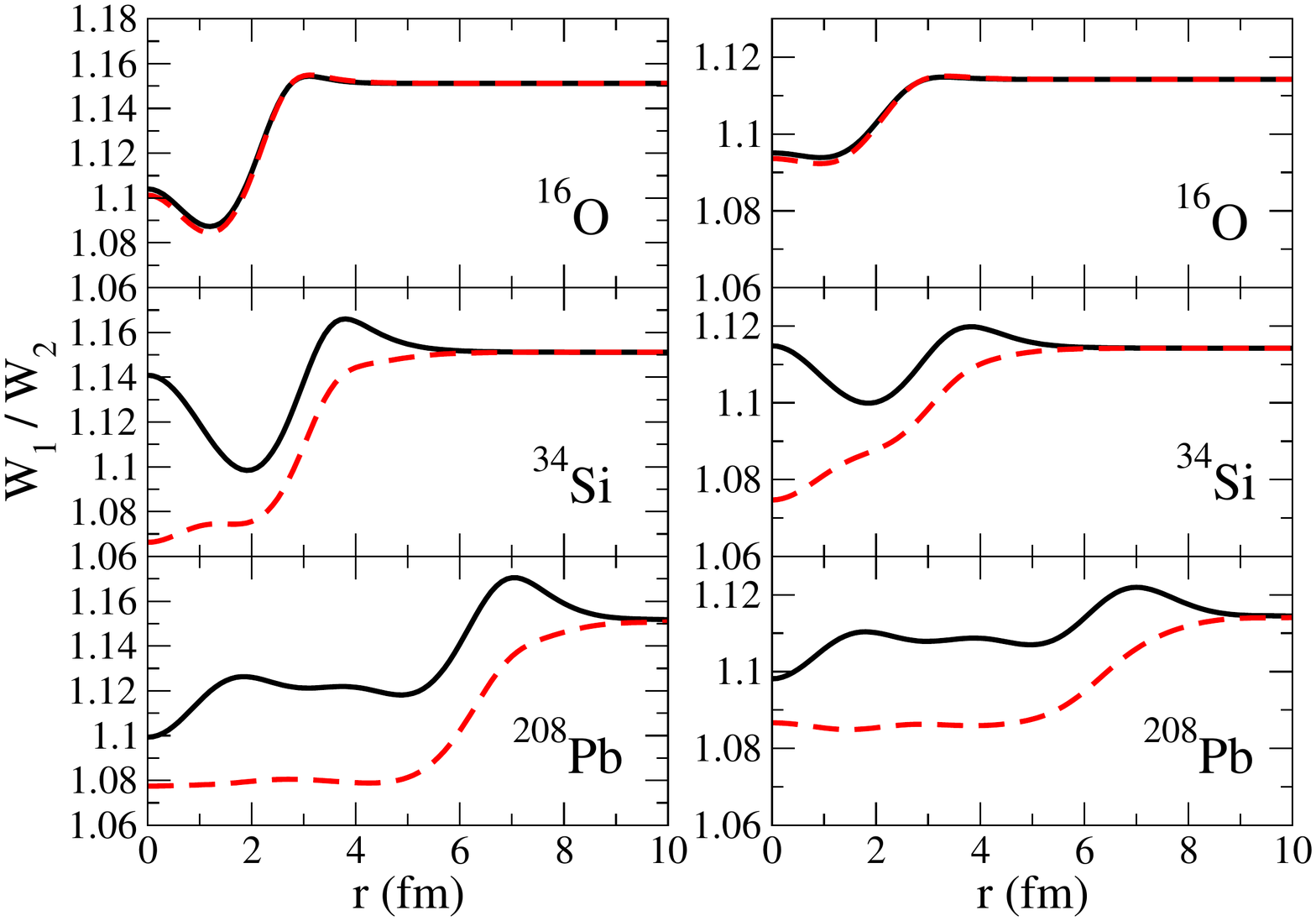}
    }
    \caption{Radial dependence of the proton (solid) and neutron (dashed) ratio $(\frac{W_1}{W_2})$ of parameters of the spin-orbit potential 
    Eq.~(\ref{eq:W1W2}), for the ground states of $^{16}$O, $^{34}$Si and $^{208}$Pb, calculated with the RMF effective interactions DD-ME2 (left) and DD-PC1 (right).}    
\label{fig:1} 
  \end{figure}

Figure \ref{fig:1} displays the radial dependence of the proton and
neutron ratio ${W_1}/{W_2}$ of parameters of the spin-orbit potential 
Eq.~(\ref{eq:W1W2}), for the self-consistent ground states of
$^{16}$O, $^{34}$Si and $^{208}$Pb, calculated with two of the most
successful RMF effective interactions DD-ME2 \cite{LNVR.05}
(meson-exchange) and DD-PC1 \cite{ddpc} (point-coupling). It should be
noted that, in contrast to the value of the ratio $W_1/W_2 = 2$ used in
standard non-relativistic Hartree-Fock calculations, in the RMF
case the ratio ${W_1}/{W_2}$ is close to one. The absolute deviation
from unity can be attributed to the contribution of the $\rho$-meson
exchange, that is, the explicit contribution of the isovector-vector
channel: in the absence of the isovector degree of freedom
in the interaction Lagrangian, Eq. (\ref{eq:W1W2}) gives $W_1=W_2$.
The isovector contribution is, of course, also responsible for the
difference between the effective proton and neutron single-particle
potentials, while the radial (density) profiles depend on the shell
structure of occupied orbitals in the self-consistent solution for a
particular nucleus. In this respect, especially interesting is the
case of $^{34}$Si, for which a possible central depletion of the
proton density distribution has been analysed using a variety of
theoretical approaches \cite{gra,yao}, and experimental constraints on
the strength of the two-body spin-orbit interaction have been reported
\cite{burg,muts}. For the effective interaction DD-ME$\delta$ that 
explicitly includes contributions from both $\rho$ and $\delta$ meson 
exchange in the direct term, the isovector channel of the spin-orbit potential 
is enhanced when compared to DD-ME2, although in both models the total isovector part of 
the spin-orbit potential is an order of magnitude weaker than the isoscalar contribution \cite{RM.11}.

\begin{figure}[]  
      {\includegraphics[width=0.8\textwidth]{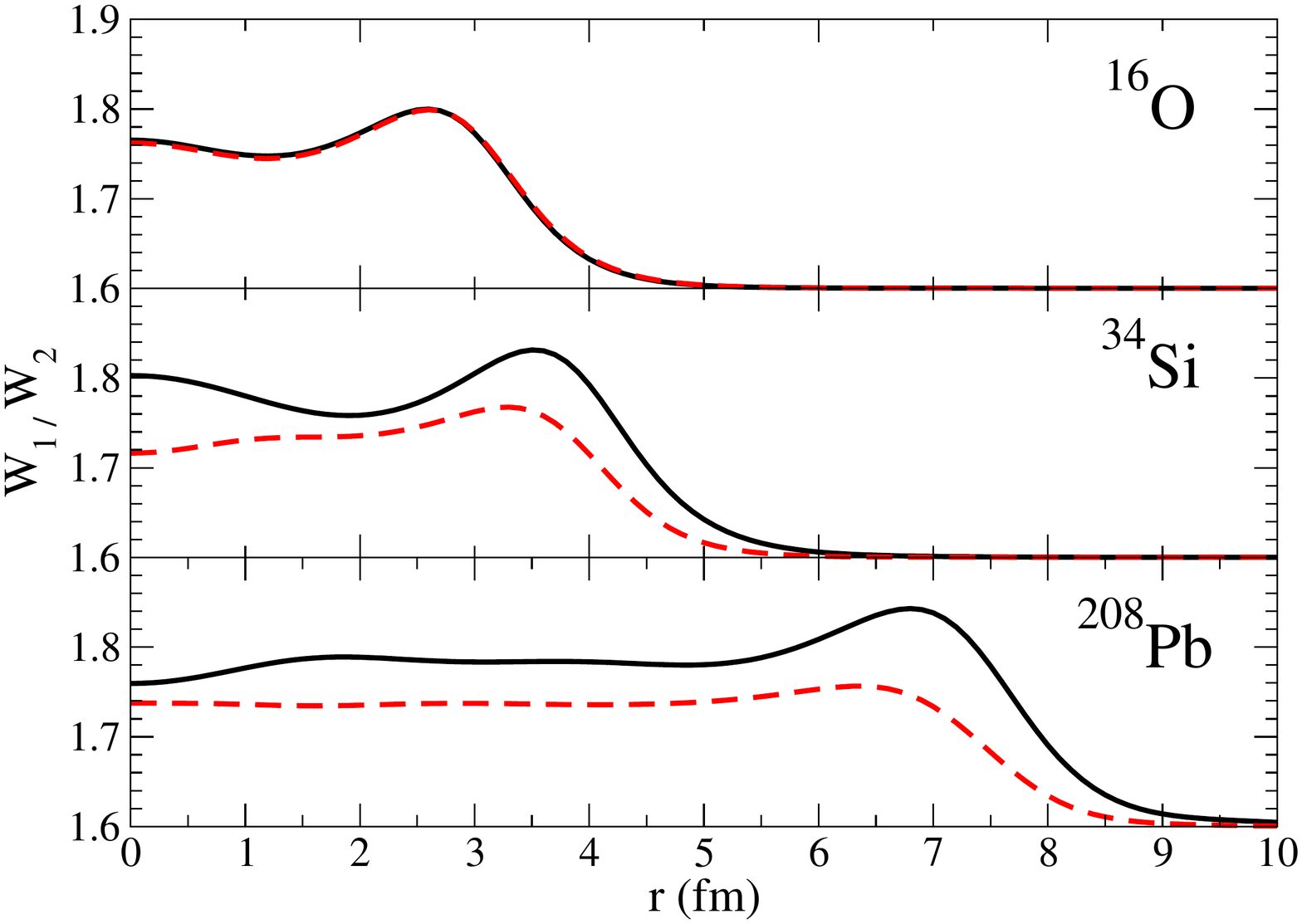}
    }
    \caption{Same as in the caption to Fig. \ref{fig:1}, but the ratio of the parameters is given by Eq.~(\ref{eq:W1W2fierz}) for the case of the relativistic Hartree-Fock effective interaction PKO2.}   
\label{fig:2} 
  \end{figure}

To illustrate the effect of the exchange terms in the RHF
approximation on the single-nucleon spin-orbit potential the radial dependence of the ratio ${W_1}/{W_2}$
for protons and neutrons for the same nuclei are plotted in Fig.
\ref{fig:2} using
Eq.~(\ref{eq:W1W2fierz}). This corresponds to using the RMF Lagrangian
(Eq.~(\ref{lag})), obtained by performing the Fierz transformation on
the interaction terms of the point-coupling RHF Lagrangian
(Eq.~(\ref{eq:lagdensity})). The effective RHF interaction is PKO2
\cite{long} which includes the $\sigma$, $\omega$, and $\rho$ meson
exchange, but not the pion or the $\delta$-meson. The most important
result is that in this case the overall value of the ratio
${W_1}/{W_2}$ is around 1.8. This is significantly larger than in the
simple RMF approach based on the Hartree approximation, and much
closer to the value 2 which characterises standard non-relativistic HF
calculations based on Skyrme forces. The difference with respect to
the latter is due to the fact that there is already an isovector
dependence of the effective spin-orbit potential for the Lagrangian
PKO2 which arises because of the $\rho$-meson exchange contribution,
and also due to the non-relativistic reduction of the single-nucleon
Dirac equation to the Schr\" odinger-like form Eq.~(\ref{Sch}) that
explicitly includes the effective spin-orbit potential. 

An effect that has been attributed to the isospin dependence of the
effective spin-orbit potential is the change (kink) of charge isotope 
shifts across the $N=126$ shell gap \cite{Ang.04,Coc.13}. The charge
isotope shift is the difference between the charge radius $< r^2_{ch}
>$ of a given isotope with respect to the reference nucleus. The best
known example is the kink in the isotope shifts of even Pb nuclei and,
more recently, a similar effect has also been observed in Polonium
isotopes \cite{Coc.13}.  Numerous calculations over the last twenty
years have shown that all relativistic mean-field effective
interactions, both at the RMF level (without or with inclusion of the
isovector scalar $\delta$ meson) and in the RHF approach, reproduce
the empirical kink in the isotope shifts of even Pb isotopes
\cite{Ring96,shar,RM.11,RF.95,par_PKO2}.  This was explained by a
relatively weak isospin dependence of the corresponding spin-orbit
potentials. Conventional Skyrme HF parameterizations with $W_1/W_2 =
2$ were unable to reproduce the kink and, therefore, in
Ref.~\cite{RF.95} the Skyrme framework was extended with an additional
degree of freedom in the spin-orbit channel which allows to modify the
value of the ratio $W_1/W_2$. This simple modification of the Skyrme
functional, in which the relative weights of the neutron and proton
contributions to the spin-orbit potential can be freely adjusted,
produces values for the isotope shifts of Pb in reasonable agreement
with data.  
\begin{figure}[]  
      {\includegraphics[width=0.6\textwidth]{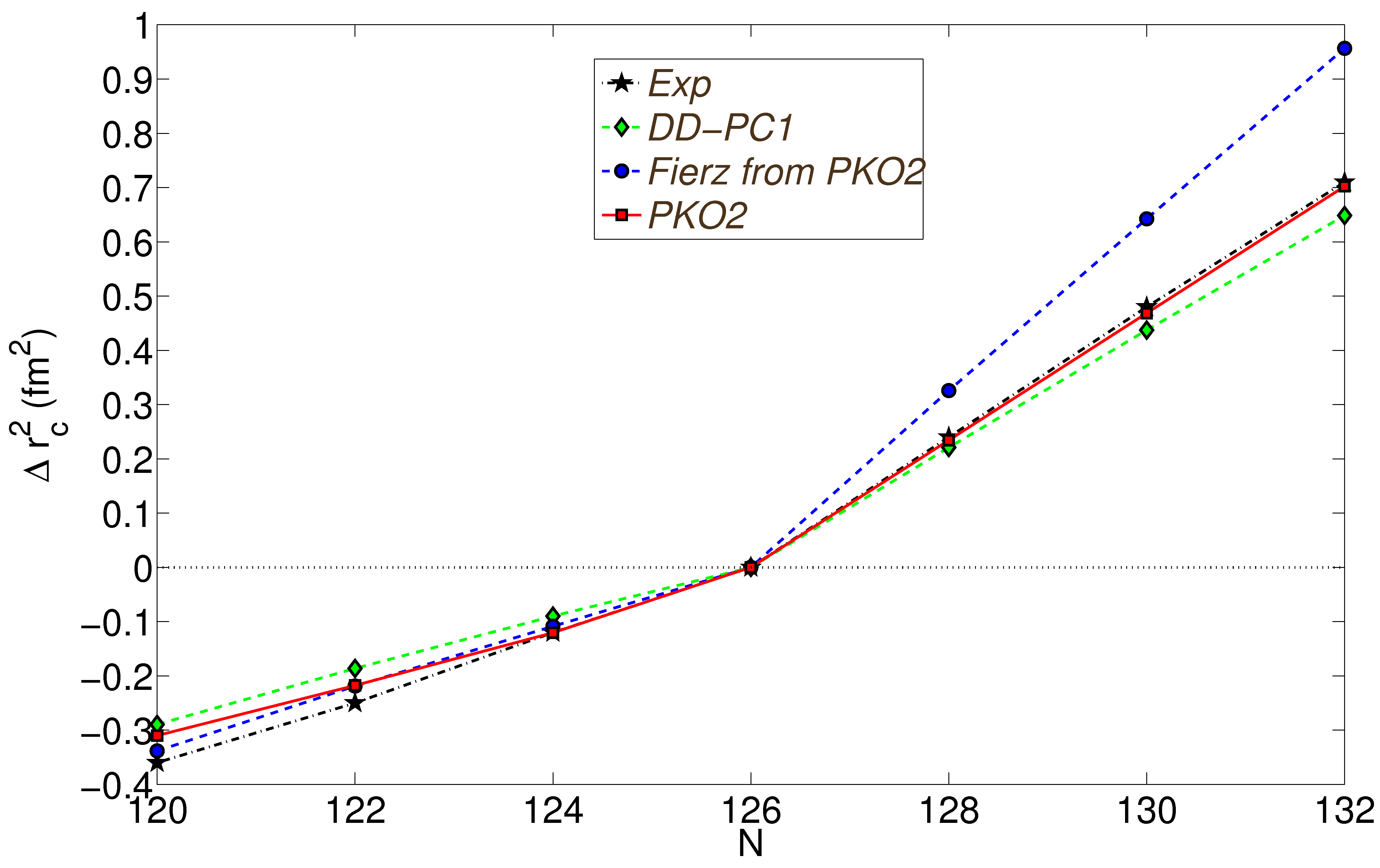}
    }
    \caption{Isotope shifts for even-A Pb nuclei with respect to the reference nucleus $^{208}$Pb. Experimental values \cite{Ang.04,Coc.13} are shown in 
    comparison with theoretical results obtained in the RMF calculation with the effective interaction DD-PC1, using the relativistic 
    Hartree-Fock effective interaction PKO2, and with the RMF model obtained by performing the Fierz transformation of 
    the point-coupling approximation of PKO2.}   
\label{fig:3} 
  \end{figure}

In Figure \ref{fig:3} we plot the experimental isotope shifts for
even-A Pb nuclei with respect to the reference nucleus $^{208}$Pb, in 
comparison with results obtained in the RMF calculation with the
effective interaction DD-PC1, using the relativistic Hartree-Fock
effective interaction PKO2, and with the RMF model obtained by
performing the Fierz transformation of the equivalent point-coupling
approximation of PKO2. In all three cases the theoretical values
reproduce the empirical kink at $N=126$ and, in particular, the kink
is most pronounced in the RMF calculation with the Fierz-transformed
effective interaction PKO2, even though ${W_1}/{W_2} \approx 1.8$ for
this model.  This result is consistent with a more recent
interpretation of the change of charge isotope shifts across the
$N=126$ shell gap \cite{GSR.13}, in which the kink is attributed to
the occupation of the $ 1 i_{11/2}$ neutron orbital and the resulting
overlap between neutron and proton orbitals with the same principal
quantum number, $n=1$. It was noted that effective forces for which
the $ 1 i_{11/2}$ neutron orbital has a significant occupation above
$N=126$, display an increase in the isotope shift of the $n=1$ proton
states. This is because when neutrons gradually occupy the $ 1
i_{11/2}$ orbital, proton states expand to larger radii to maximally
overlap with the additional neutrons \cite{GSR.13}. In the present
calculation, both for DD-PC1 and PKO2, the neutron orbitals $ 1
i_{11/2}$ and $ 2 g_{9/2}$ are almost degenerate above $N=126$, and
this leads to significant occupation of $ 1 i_{11/2}$ and the
resulting sudden increase in the isotope shifts. In fact, the
quasi-degeneracy of $ 2 g_{9/2}$ and $ 1 i_{11/2}$ corresponds to an
approximate realization of pseudospin symmetry of single-nucleon
states with $(n, l, j=l+1/2)$ and $(n-1, l+2, j=l+3/2)$.  When the
Fierz transformation is performed on the point-coupling approximation
of PKO2, the equivalent RMF Lagrangian leads to the lowering of the
orbital $ 1 i_{11/2}$ below $ 2 g_{9/2}$. Although it cannot directly
be compared to data \cite{duguet}, this discrepancy with the
experimental spectra of $^{209}$Pb and $^{211}$Pb is probably caused
by the fact that the parameters of the equivalent Lagrangian are not
fine-tuned after performing the point-coupling approximation.
Nevertheless, it leads to the pronounced kink shown in
Fig.~\ref{fig:3}. In fact we note that the best agreement with the
empirical kink is obtained with those interactions for which
pseudospin symmetry is realized in the single-neutron spectra (here
DD-PC1 and PKO2, but also other relativistic interactions). If this
symmetry is broken by further lowering $ 1 i_{11/2}$, below $ 2
g_{9/2}$, the kink in the isotope shifts becomes too strong compared
to data (cf. Fig.~\ref{fig:3}). The presence of the kink in the
isotope shifts and the relativistic models that reproduce the data
thus provide evidence for the occurrence of pseudospin symmetry in
neutron-rich Pb nuclei.

In conclusion, we have analyzed the isospin dependence of the
effective spin-orbit potential in standard relativistic meson-exchange
or point-coupling (contact) effective interactions, when used in the
mean-field (Hartree) or Hartree-Fock approximations. By performing a
non-relativistic reduction of the single-nucleon Dirac equation to a
Schr\" odinger-like form that explicitly exhibits the spin-orbit
potential, the corresponding isospin dependence can be directly
compared to that of the non-relativistic Hartree-Fock models based on
effective Skyrme forces. This isospin dependence can be characterised
by the ratio ${W_1}/{W_2}$ of the two parameters in the expression for
the effective spin-orbit potential Eq.~(\ref{W1andW2}). In
conventional non-relativistic Hartree-Fock mean-field calculations
based on the Skyrme force $W_1/W_2 = 2$, whereas in standard RMF
models this ratio is close to 1. The deviation from 1 arises because
of the explicit isovector contribution to the spin-orbit potential. 
It should be noted that, because of the medium-dependence of the
effective coupling parameters, either through an explicit density
dependence or higher-order self-interaction terms, the ratio
${W_1}/{W_2}$ is density dependent in the relativistic approach. In
the case of relativistic Hartree-Fock models, to evaluate the effect
of exchange terms we have performed a Fierz transformation of the
point-coupling RHF Lagrangian and derived an equivalent RMF Lagrangian
that, in addition to the isovector-vector contribution of the original
RHF Lagrangian ($\rho$-meson exchange), contains also an
isovector-scalar term. As a result, the ratio ${W_1}/{W_2} \approx
1.8$ is much closer to the value that characterises standard Skyrme
Hartree-Fock models. This result is important in view of recent
experimental efforts to explore the isospin dependence of spin-orbit
forces in nuclei. It shows that, when comparing with results obtained
using conventional Skyrme HF models, Fock terms should also be treated
explicitly in relativistic mean-field models or, if one wants to
preserve the advantage of local density functionals, the
isovector-scalar channel has to be taken into account in addition to
the usual isovector-vector contribution. Of course, this channel has
been considered before in relativistic structure models, however the
standard data (masses, radii) could not be used to discern between the
two isovector channels. Information on the isospin dependence of the
energy spacings between spin-orbit partner states could thus be used
to determine the isovector-scalar channel contribution. We have also
shown that the reproduction of the empirical kink in the isotope
shifts of even Pb nuclei by relativistic effective interactions points
to the occurrence of pseudospin symmetry in the single-neutron spectra
in these nuclei.

\section*{Acknowledgement}
The authors thank T. Nik\v{s}i\'{c} and O. Sorlin for fruitful discussions. This work has been supported in part by the QuantiXLie Centre of Excellence.

\end{document}